# Nanostructure of edge dislocations in a smectic C* liquid crystal

C. Zhang[1], A.M. Grubb, A.J. Seed[2], P. Sampson, A. Jákli[1], O.D. Lavrentovich[1]

[1]Liquid Crystal Institutes, Kent State University, Kent, OH 44242, US

[2]Department of Chemistry and Biochemistry, Kent State University, Kent, OH 44242, US

*Abstract:*

*We report on the first direct nanoscale imaging of elementary edge dislocations in a thermotropic smectic C\* liquid crystal with the Burgers vector equal to one smectic layer spacing d. We find two different types of dislocation profiles. In the dislocation of type A, the layers deformations lack mirror symmetry with respect to the plane perpendicular to the Burgers vector; the dislocation core size is on the order of d. In the dislocation of type S, the core is strongly anisotropic, extending along the Burgers vector over distances much larger (by a factor of 4) than d. The difference is attributed to a different orientation of the molecular tilt plane with respect to the dislocation's axis; the asymmetric layers distortions are observed when the molecular tilt plane is perpendicular to the axis and the split S core is observed when the molecules are tilted along the line.*

Linear defects in materials with broken translational symmetry, called dislocations, determine many static and dynamic properties of these materials [1]. The structure and behavior of dislocations is relatively well studied for the case of regular solids, especially metals [2]. Dislocations in soft matter, such as smectic liquid crystals and block copolymers, play a similarly important role [1] [3], as evidenced by the studies of rheological effects [4–8]. In some cases, such as the vicinity of a smectic A (SmA)-to smectic C (SmC) phase transition, accompanied by a tilt of molecules within the layers, the presence of dislocations can be verified directly by optical microscopy [9]. One of the important questions that remains unanswered is the detailed structure of the dislocation core, i.e., the region at the "center" of the defect where the deformations are too strong to sustain the usual type of order. The spatial extension of the core is of the order of a few characteristic periods of the positional order and in most smectics it is in the range of 1-10 nanometers, which calls for imaging techniques with resolution much higher than the optical one. An obvious solution is to use electron microscopy, but the latter is often limited by the soft nature of smectics and by the need to align the material. For example, to observe an edge dislocation, in which the Burgers vector **b** is perpendicular to the defect axis, a desirable orientation of the smectic layers is parallel to the probing beam. The latter is very difficult to achieve, as the smectic layers tend to be parallel to the bounding plates/interfaces. Because of all these difficulties, the nanometer-resolved images of edge dislocations in smectics are available only for very few materials, such as the lyotropic lamellar phase of phospholipids [10] and bent-core thermotropic smectics [11].

In this work, using cryo-transmission electron microscopy (cryo-TEM), we present the first direct observation of elementary edge dislocations in a thermotropic smectic C* (SmC*) phase formed by rod-like chiral molecules. The study reveals two different types of profiles of an edge dislocation with the Burgers vector $b = d$, where $d$ is the smectic periodicity. In the first type, the layer deformations lack mirror symmetry with respect to the plane of the extra layer. The asymmetric structure is called an A-core dislocation. In the second type, the core is strongly anisotropic, extending along the Burgers vector **b** over the distances $2\xi_z$ much larger (by a factor of 4) than the core size $2\xi_x$ measured in the direction perpendicular to the Burgers vector. This type represents a split core and is called an S-type. The observation of S-cores confirms a long-standing prediction by Allen and Kleman [12]. We suggest that the distinct A and S cores are caused by the different direction of molecular tilt within the smectic layers, which is perpendicular to the dislocation's axis in the A case and parallel to it in the S case.

For our experiments we used (*S*)-4-(1-methylheptyloxy)phenyl 4-(2- dodecyloxy-1,3-thiazol-5-yl)benzoate, abbreviated as AG14 in what follows [11]. In cooling it exhibits the following phase sequence: Iso 104ºC SmA 102ºC SmC* 72ºC Cr. Small angle X-ray scattering studies show a single peak at $q = 0.18 \text{Å}^{-1}$ in the entire SmC* range corresponding to layer spacing of $d_X = 34.9 \text{Å}$, which is smaller than the fully stretched molecular length of $l = 39$ Å suggesting a tilt angle of $\theta_X = 23.5°$. The normal $\hat{\mathbf{v}}$ to the SmC* layers is also the axis of the heliconical director $\hat{\mathbf{n}} = \{\sin\theta_0 \cos\varphi, \sin\theta_0 \sin\varphi, \cos\theta_0\}$, specifying the local molecular orientation that twists in space around $\hat{\mathbf{v}}$; here $\theta_0$ is the polar angle that the molecules make with the normal to the layers, $\varphi = \hat{q}z$ is the azimuthal direction of the molecular tilt, $\hat{q} = 2\pi/P$ is the twist wavevector, and $P = 20\,\mu\text{m}$ is the heliconical pitch [11]. The latter is much larger than the smectic spacing.

Tilt angle measurements were carried out by applying square wave electric fields and measuring the angular difference between two directions of the optical axes corresponding to the positive and negative voltages, found $\theta_{opt} = 22°\pm2°$ for the whole temperature range. The closeness of $\theta_X$ and $\theta_{opt}$ indicates that the titled molecules are fairly straight in the SmC* phase. The material has a low viscosity due to the compact nature of the 1,3-thiazole ring, and forms chevron-free structures in 5-10µm thick films. These properties are not only advantageous for display purposes, but are also very useful in our TEM studies. When the material is sandwiched between two plasma-treated continuous carbon films, is shows a "bookshelf" alignment of layers that are parallel to the probing electron beam. AG14 turned out to be the only material out of dozen smectics explored by us that yielded this alignment and thus allowed us to explore the fine structure of edge dislocations.

Cryo-TEM measurements were carried out on a FEI Tecnai F20 microscope operating at 200 kV. A Gatan cryo-holder (model 626.DH) made it possible to keep the specimen temperature below -170°C throughout the TEM observation. All images were recorded using a Gatan 4K Ultra Scan CCD camera. The films were heated to the isotropic phase and cooled to the desired temperatures, then rapidly quenched to liquid nitrogen (–196°C) to preserve the liquid crystalline structures [13].

The films were normally previewed rapidly at a dose of 20 e$^-$/nm$^2$; selected areas were then imaged at a dose level of 200e/nm$^2$, which we found did not cause any radiation damage. Previous studies of ~100 nm thermotropic bent-core liquid crystals films carried out with this instrument visualized smectic layers with resolution better than 0.7 nm. [13–16]. The contrast in cryo-TEM image, Fig.1, is due to the difference in electron density of the aromatic core and the hydrocarbon tail. Lighter image areas in Fig.1 correspond to regions filled with hydrocarbon tails with lower electron absorption. To maintain the contrast through the entire thickness of the sample, the layers have to be aligned parallel to the electron beam, with angular deviation less than $\alpha < \tan^{-1}(d/L)$. For $d \approx 4$ nm and for the film thickness $L \sim 100$ nm, this means $\alpha < 3°$. The condition also implies that the periodicity $d$ measured from the TEM images differs from the actual periodicity by less than 0.1%.

A typical TEM image of thin ($L \sim 100$ nm) samples quenched from the SmC* phase at 98°C shows a periodic transmitted electron intensity profile (Figure 1a), indicating uniform alignment of smectic layers in the so-called bookshelf geometry over 0.1µm$^2$ areas. The fast

Fourier transform (FFT) pattern [Figure 1 (b)] obtained from a 300 nm x 350 nm area reveals a periodicity $d = 3.83$ nm, which is very close to $l = 3.9$ nm and definitely larger than the layer spacing 34.9Å measured by X-ray in bulk. This shows that the substrate suppresses the director tilt angle to a value $\theta_o = \cos^{-1}(d/l) = \cos^{-1}(3.83/3.9) = 13^o$ or even less; the same effect has been observed for other tilted smectics. [14]

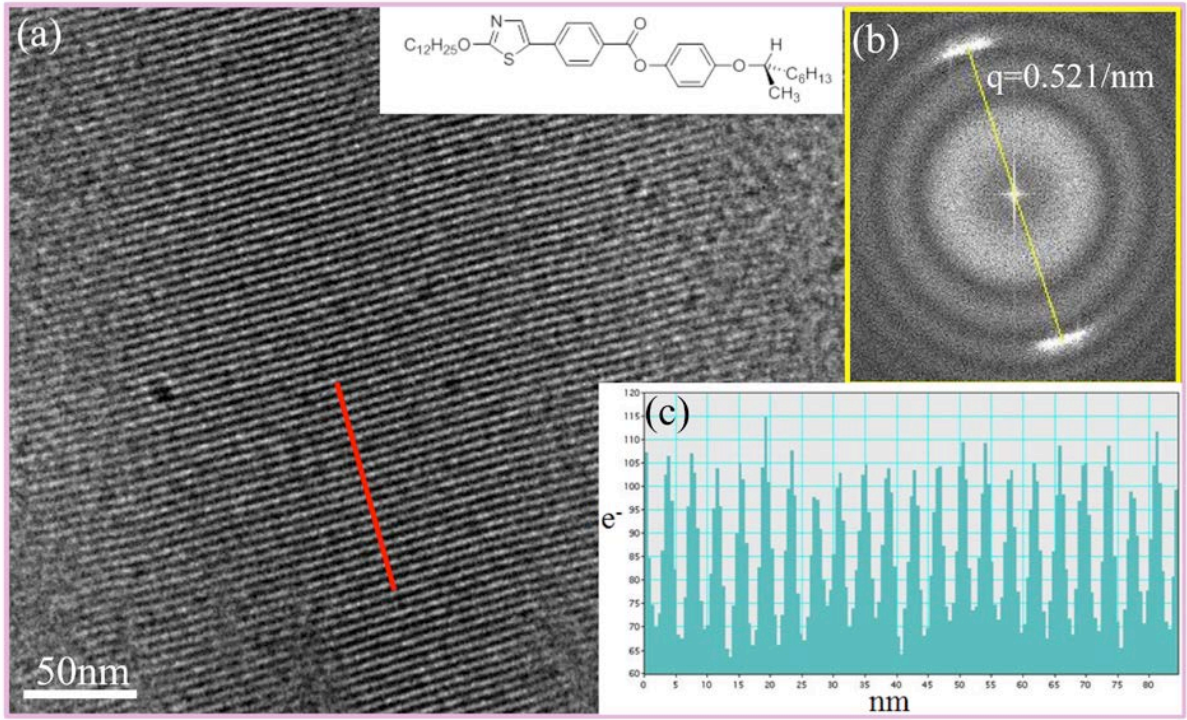

*Figure 1: Typical cryo-TEM image of the studied material quenched from 98°C. (a) 300 nm x 350 nm area showing uniform smectic layers normal to the substrate. Inset shows the molecular structure; (b) The FFT image of the whole area showing the periodicity (2/0.521) nm=38.3 Å. (c) The intensity profile along the red line marked in (a).*

Apart from large uniform areas, a number of elementary edge dislocations with Burgers vector $b = d$ are also observed. These come in two different shapes that we label as A and S edge dislocations, see Figs. 2 and 3, respectively. To facilitate the discussion, we define two planes. One is the glide plane (GP) formed by the dislocation axis along the $y$-axis, and its Burgers vector **b** along the $z$-axis. In our coordinates, GP is the $x=0$ plane. In all experiments, the GP is perpendicular to the plane $xz$ of view, which enables the detailed study of the core structure. The second plane is the molecular tilt plane (MTP), determined by the local director $\hat{\mathbf{n}}$ and the layer normal $\hat{\mathbf{v}}$.

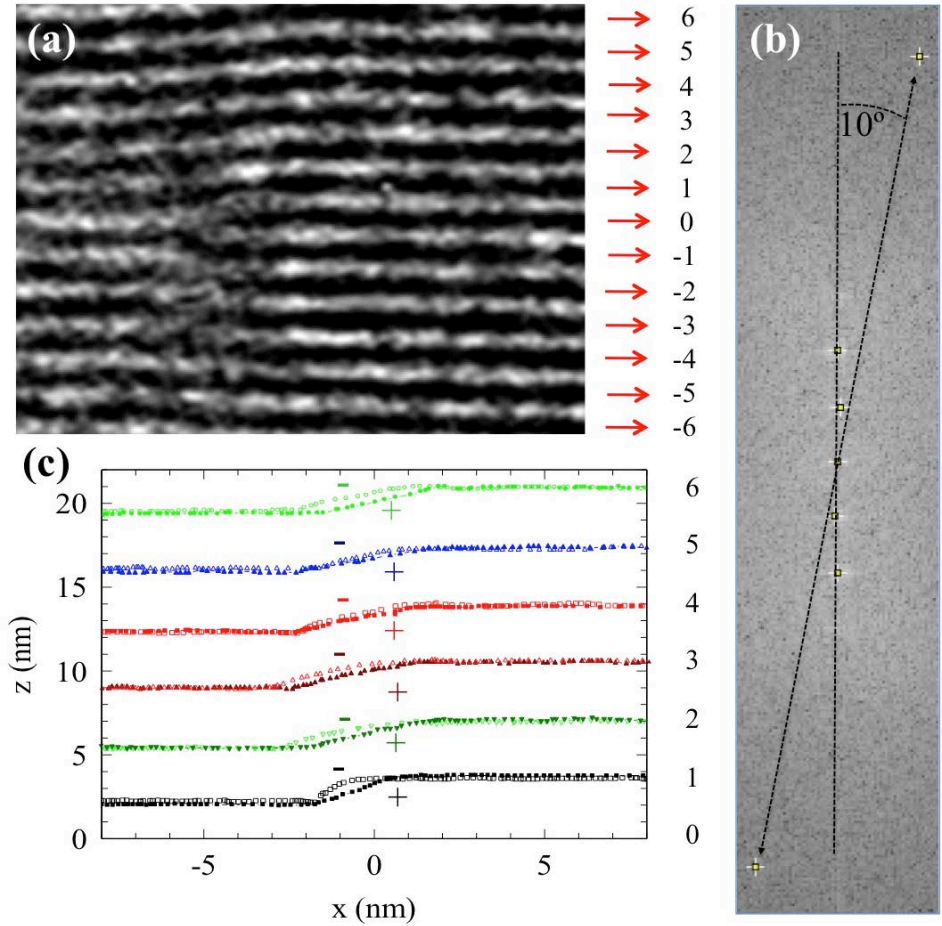

*Figure 2: Asymmetric-core A edge dislocation with the Burgers vector b=d. (a) TEM image; (b) FFT of the image in (a) showing the layered structure with d=3.83 nm periodicity (peaks along the vertical line) and peaks along the line tilted by 10º from the vertical, corresponding to 0.55 nm periodicity. (c) The x-z dependence of the layer shifts $|u(x,z)|$ around the core; the symbol "-" labels the layers located at z<0; "+" labels the layers from the upper half.*

The A type edge dislocation is shown in Figure 2. The TEM texture shows one extra smectic layer (dark band labeled with "0") inserted on the right hand side, Fig. 2a. The core region, in which this layer ends, Fig.2c, is relatively "isotropic", extending along the $x$- and $z$-axes by the distances $2\xi_x$ and $2\xi_z$, respectively, which are approximately equal to each other and are on the order of the smectic spacing $d$. The most notable feature is that the displacements of smectic layers above (layers labeled with "+" in Fig.2c) and below ("-"labels) the plane $z=0$ are not symmetric, $u(x,z) \neq -u(x,-z)$. In particular, the tilt of layers $\partial u / \partial x$ with respect to the $x$-axis, is larger for $z<0$ than for $z>0$, Fig.2c. This is in sharp contrast with the symmetric edge dislocation profiles observed in cholesterics [17] and bent-core smectics A [14] and predicted by the theory for one-dimensionally periodic materials with no molecular tilt within the layers, such as the SmA [18] [19] [20].

The A type dislocation in Fig.2a is surrounded by layers in which the molecules tilt in the direction perpendicular to the dislocation line. MTP is perpendicular to GP, being either parallel to the $xz$ plane of observation or close to it. The FFT pattern in Fig.2b, corresponding to the real space image in Fig.2a, exhibits two sets of reflexes. The first set contains vertically spaced peaks associated with the layer periodicity $d = 3.83$ nm along the $z$-axis. The second set is represented by two peaks located at the line that makes an angle about 10º with the $z$ axis. The angle 10º agrees well with the estimate of the molecular tilt within the layers. The $q$ value of the tilted peak infers a periodicity of 0.55 nm, which corresponds well to the distance between the thiazole and benzene ring (see inset in Fig.1a) along the molecule. Therefore, the MTP is perpendicular to the GP of the dislocation in Fig.2a.

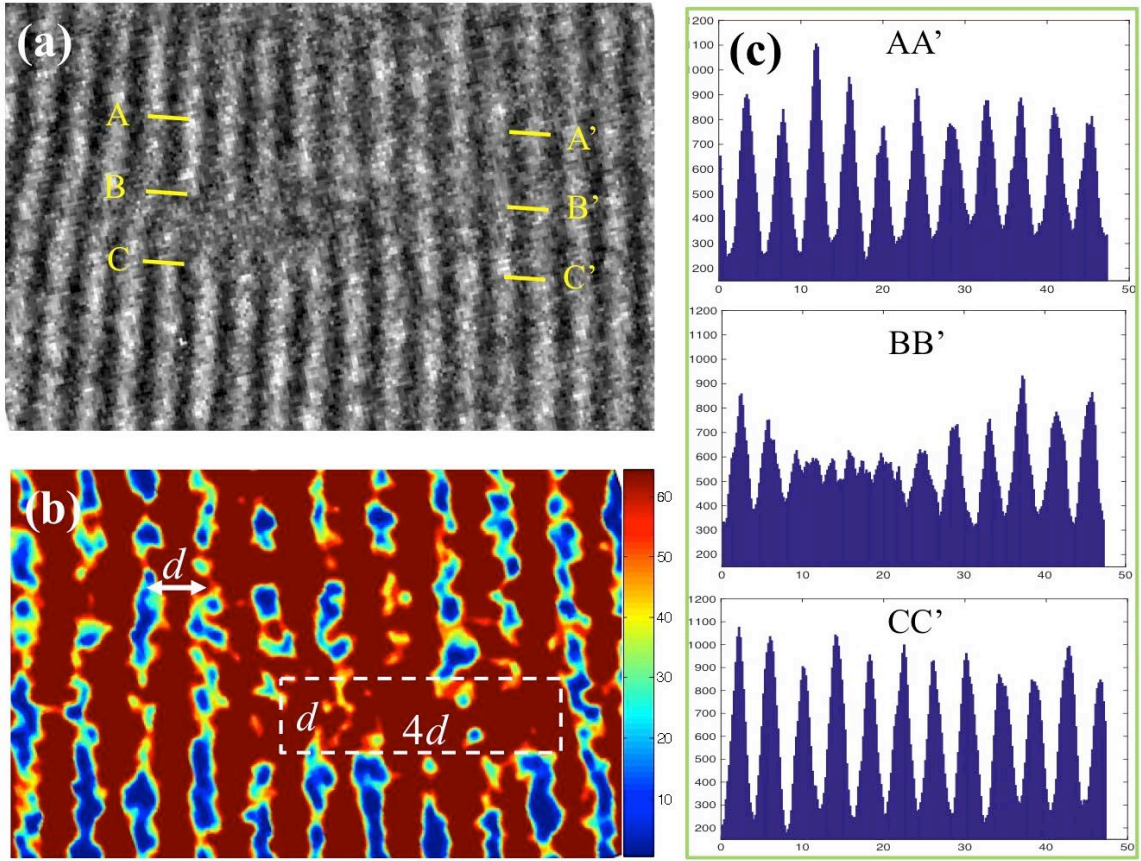

*Figure 3: Split-core S edge dislocations. (a) Grey-scale TEM image of an area with split-core edge dislocation. (b) A color-enhanced TEM image of another area with split-core edge dislocation. Dotted rectangle covers the split core with the aspect ratio 4:1. (c) Transmitted electron density profiles measured along the lines AA', BB' and CC' shown in part (a). Note the periodic nature of density variation along the lines AA' and CC' and reduction in the amplitude of modulations along the BB' line over a ~15 nm segment.*

A very different S type of edge dislocation core is presented in Fig.3. In this case, there are no clear FFT reflexes that could be associated with the molecular tilt in the $xz$ plane of observation, suggesting that the molecules are tilted along the axis of dislocation; MTP is parallel to GP. The S core is highly anisotropic, extending along the Burgers vector **b** over a distance $2\xi_z \approx 4d$ that is much larger that the core extension $2\xi_x \approx d$ measured in the direction perpendicular to the GP, Fig.3a. The anisotropic core involves multiple layers, $n>1$, that are being disrupted and "melted" into a nematic-like region. In Fig.3a,bc, there are $n=4$ layers on one side of the GP and 5 layers on the other side that are clearly disconnected. Figure 3c shows three profiles of the transmitted electron intensity measured above (AA'), below (CC') and in the middle (BB') of the core region, demonstrating that the material density within the core (along the cut BB') is practically constant over a large distance on the order $n=4$ smectic periods, $2\xi_z \approx 4d$; the latter implies melting of the smectic positional order. Note that the average transmitted intensity is the same inside and outside the defect area, indicating that the average density of molecular packing is practically the same. This behavior is different from the case of bent-core smectics [14], in which the core of dislocations is packed less densely than the uniform areas.

The experimental results in Figs. 2 and 3 demonstrate that the nanoscale structure of an elementary edge dislocation depends on the angle $\varphi$ between MTP and GP, which is close to either $\pi/2$ (type A dislocation) or 0 (S dislocation), Fig.4. Cryo-TEM textures do not allow us to measure $\varphi$ accurately; the conclusions that $\varphi = \pi/2$ in Fig.2a and $\varphi = 0$ in Fig.3a are based on whether the additional reflexes in FFT images are observed or not. In an ideal unbounded SmC* sample, the angle $\varphi = \hat{q}z$ continuously changes along the $z$-axis. However, since the pitch is very large, $P = 20\,\mu m$, the TEM textures in Figs.2a and 3a correspond to a practically constant $\varphi$; with the field of view $\Delta z$ being only 50 nm - 90 nm, the maximum variation $\Delta \varphi$ across $\Delta z$ is very small, less than $2^o$. In bounded samples, surface anchoring of the director can partially or completely suppress this rotation and favor some selected values of $\varphi$. The states with $\varphi = \pi/2$ correspond to the tangential alignment of molecules at the substrates, while for $\varphi = 0$, there is a small surface tilt, comparable to $\theta_0 \sim 10^0$.

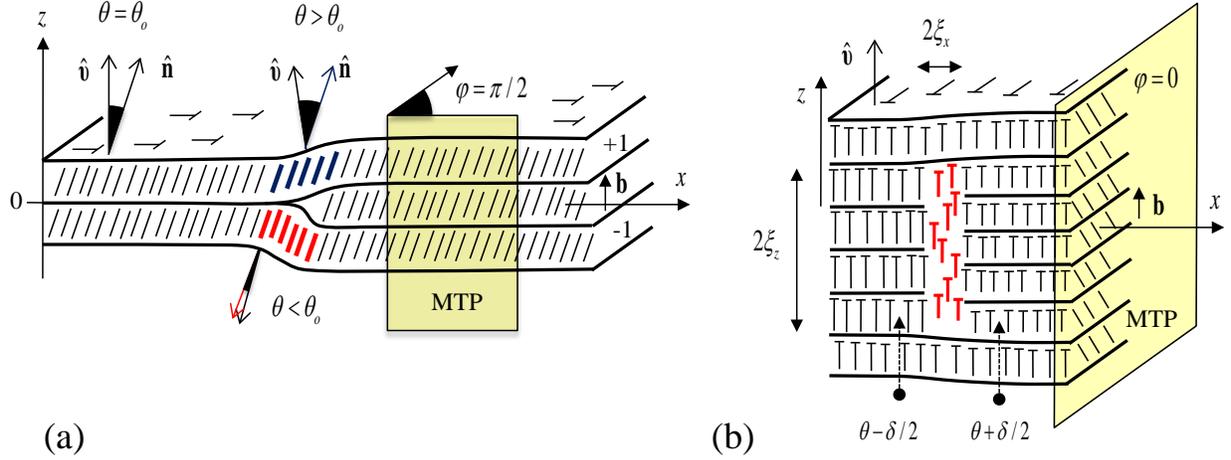

*Figure 4: Schematic images of A (a) and S (b) types of elementary edge dislocations in SmC\*. The nails show molecules tilted with respect to the plane of view; the heads are closer to the viewer than the ends. The layers tilt in (a) creates a different polar angle of the molecular tilt $\theta$ above and below the plane $z=0$. Within the S-core, the predominant director deformation is twist. Note the different direction of azimuthal tilt with respect to the glide plane of the dislocation, $\varphi = \pi/2$ in (a) and $\varphi = 0$ in (b).*

The distinctive feature of the A-dislocation is an asymmetric nature of layers displacements above and below the plane $z=0$, Fig.2c and 4a. The effect can be related to the elastic coupling between the layer deformations and the local molecular tilt $\theta$ that is different from the equilibrium value $\theta_0$ in a uniform sample. As schematically shown in Fig.4a in an approximation of a uniform director, the tilt $\partial u/\partial x$ of layers around the A dislocation imposes a lager molecular tilt $\theta > \theta_0$ for $z>0$ and a smaller tilt $\theta < \theta_0$ for $z<0$. The energy density of the distorted SmC\* with a locally fixed $\varphi$ can be expressed through the layers deformations and the local molecular tilt $\theta$, as (see, for example, [21]) $f_c = B\left[\partial u/\partial x - \tfrac{1}{2}(\partial u/\partial x)^2 + \tfrac{1}{2}\theta^2\right]^2/2 + K\left(\partial^2 u/\partial x^2\right)^2/2$. Here $B$ is the Young's modulus and $K$ is the curvature modulus. Since the energy density $f_c$ should be about the same in the $z>0$ and $z<0$ semi-planes, different local tilt of molecules can produce different profiles of layers $u(x,z>0)$ and $u(x,z<0)$ in the upper and lower semi-planes, as indeed observed in the experiment, Fig.2c. Interestingly, in the pioneering work on dislocations in wedge samples near the SmA-SmC phase transition, Meyer, Stebler and Lagerwall [9] stressed that the optical features are consistent with $\varphi = \pi/2$. However, the asymmetric character of the layer displacement could not be verified experimentally since the dislocations were viewed along the direction parallel to **b**.

In the S core, the predominant type of director deformations in the nematic-like core is the twist that allows one to accommodate a different thickness of the layers on both sides of the GP, Fig.4b. The anisotropic nature of the S core can be connected to the elastic properties of the medium by calculating the line tension of dislocation [12] [22], neglecting nonlinear effects [19] [23]. In the linear model, displacements of layers around the edge dislocation are described as $u(x,z) = -\frac{b}{4}\mathrm{sgn}(z)\left[1+\mathrm{erf}\left(\frac{x}{2\sqrt{\lambda z}}\right)\right]$, where $\lambda = \sqrt{K/B}$. [18] The line tension $F$ is then calculated by integrating the free energy density $f_c = \frac{1}{2}B\left(\frac{\partial u}{\partial z}\right)^2 + \frac{1}{2}K\left(\frac{\partial^2 u}{\partial x^2}\right)^2$ over the $xz$ plane, excluding a rectangular core area $|x| \leq \xi_x, |z| \leq \xi_z$ where the deformations are high. The result [22], $F = \frac{Kb^2}{3\pi\xi_x\lambda} + F_c$, contains the core energy $F_c$ of distortions within this rectangle. We present the core energy as $F_c = 2\xi_z\sigma_z$, where the energy density $\sigma_z$ is associated with the director twist, from $\theta_0 + \delta/2$ to $\theta_0 - \delta/2$, as one crosses the core along the $x$-axis, Fig.4b. The twist angle $\delta$ is determined by the number $n$ of layers involved in the discontinuity at the core. To estimate the relationship, we equate the total thickness $(n+1)d'$ of the layers on one side of the core to the similar quantity $nd''$ on the other side; here $d' \approx l\cos(\theta_0 + \delta/2)$ and $d'' \approx l\cos(\theta_0 - \delta/2)$. Then it is easy to see that $\delta \approx \cot\theta_0 / n$. As a result, the core energy $F_c = 2K\left(\dfrac{\cot\theta_0}{2n\xi_x} - \hat{q}\right)^2 \xi_x \xi_z \approx \dfrac{1}{2n} K\cot^2\theta_0$ decreases as $n$ becomes larger; the latter explains the tendency of the core to split along the $z$-axis. In the last expression, $\hat{q} \sim 0.3/\mu m$ is neglected since it is much smaller than $1/n\xi_x$. Using the estimates $\xi_x = d/2$, $\xi_z = nd/2$ that follow from the experimental data in Fig.3, one can rewrite the line tension of dislocation as $F = \dfrac{16nK}{3\pi} + \dfrac{K\cot^2\theta_0}{2n}$. Minimization with respect to $n$ yields $n = \sqrt{\dfrac{3\pi}{32}}\cot\theta_0$. The latter estimate correlates well with the experimental data, as for $\theta_0 \sim 10^0$ the formula predicts $n \approx 4$.

To conclude, we presented the first experimental observations of the nanoscale details of elementary edge dislocations in a weakly twisted smectic C*. We found two types of dislocations, an A type with an asymmetric profile of layers in the top and bottom semi-planes, and an S split type with the core strongly elongated along the Burgers vector and involving more than one melted smectic layer. We connect the observed features to the elastic coupling between the layer distortions and molecular polar and azimuthal tilts within the layers. In the A dislocations, the molecules tilt in the direction perpendicular to the dislocation's axis, while in the S dislocations, the molecular tilts are parallel to the defect's axis. The experimental observations pose a challenging problem of incorporating the polar and azimuthal components of the molecular tilts into the theoretical models of edge dislocations, especially at the core where the vanishing smectic order varies in space and couples to the orientational degrees of freedom; to the best of our knowledge, this problem has not been treated so far.

**Acknowledgement:** This work was supported by DMR-1410378 and DMR-1307674. The TEM data were obtained at the cryo-TEM facility at the Liquid Crystal Institute, Kent State University, supported by the Ohio Research Scholars Program "Research Cluster on Surfaces in Advanced Materials" with the help of Dr. M. Gao. We thank R. Kamien for discussions and M. Salili for the help with the data.